# Perfect matching of reactive loads through complex frequencies: from circuital analysis to experiments

Angelica V. Marini, *Student Member, IEEE*, Davide Ramaccia, *Senior Member, IEEE*, Alessandro Toscano, *Senior Member, IEEE* and Filiberto Bilotti, *Fellow, IEEE*

*Abstract*—The experimental evidence of purely reactive loads impedance matching is here provided by exploiting the special scattering response under complex excitations. The study starts with a theoretical analysis of the reflection properties of an arbitrary reactive load and identifies the proper excitation able to transform the purely reactive load into a virtual resistive load during the time the signal is applied. To minimize reflections between the load and the transmission line, the excitation must have a complex frequency, leading to a propagating signal with a tailored temporal envelope. The aim of this work is to design and, for the first time, experimentally demonstrate this anomalous scattering behavior in microwave circuits, showing that the time-modulated signals can be exploited as a new degree of freedom for achieving impedance matching without introducing neither a matching network nor resistive elements, that are typically used for ensuring power dissipation and, thus, zero reflection. The proposed matching strategy does not alter the reactive load that is still lossless, enabling an anomalous termination condition where the energy is not dissipated nor reflected, but indefinitely accumulated in the reactive load. The stored energy leaks out the load as soon as the applied signal changes or stops.

*Index Terms*— Complex frequency, Impedance matching, Time modulation, Virtual perfect absorption, Zero scattering

## I. INTRODUCTION

IMPEDANCE MATCHING is one of the fundamental concepts in microwave circuit theory, allowing the maximum energy transfer to a load by cancelling reflections at its terminals [1], [2]. Impedance matching is invoked when the impedances of the feeding transmission line $Z_0$ and of the load $Z_L$ do not satisfy the condition $Z_L = Z_0$, provided that the generator internal impedance equals to the input impedance of the feeding transmission line connected to the load. Without impedance matching, reflections occur inevitably at the terminals where the impedance discontinuity takes place for ensuring the continuity of the physical quantities involved in the signal propagation, *i.e.*, voltages and currents, along the transmission line. To avoid such undesired reflections, in the last century, several impedance matching techniques have been developed, which are based on the concept of impedance transformation, *i.e.*, adding a properly designed microwave network composed by lumped and distributed elements between the line and the load (Fig.1(a)). A number of design techniques of matching networks [3]–[6] have been proposed in the state of the art to reach the theoretical limits imposed by Bode-Fano theorem [1] to achieve the best impedance matching conditions for antennas [7]–[10], power amplifiers [11], [12] and energy transfer applications [13]–[15], spacing from adapting matching [16]–[18] to genetic algorithms techniques [19], [20]. Indeed, considering a frequency-dependent complex load $Z_L(\omega) = R(\omega) + jX(\omega)$ fed by a lossless transmission line with a purely resistive characteristic impedance $Z_0 = R_0$, the matching network is designed such that the load reactance vanishes and the load resistance transforms into $R_0$ at the desired frequency $\omega_0$. Therefore, conventional impedance matching techniques act on the network topology and the involved circuital elements, *i.e.*, capacitance, inductance, and resistance (Fig.1(a)).

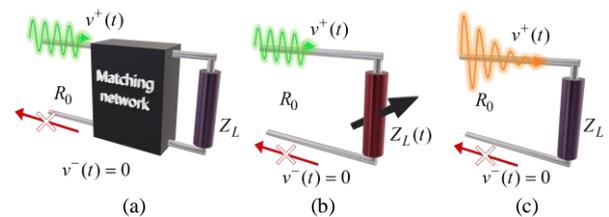

Fig. 1. Schematic representation of the impedance matching techniques: impedance transformation through (a) matching network, (b) temporal modulation of load, (c) temporal modulation of the excitation signal.

Recently, a new degree of freedom has been introduced for achieving perfect matching: the temporal modulation of the constituting elements of the complex load [21]. Despite the earlier studies dated back to the 1960s, time-varying elements in electronic circuits [22], [23], as well as space-time-varying material properties for engineering wave propagation [24]–[31], have only recently attracted the attention of researchers, thanks to the possibility to achieve magnet-less non-reciprocity and frequency conversion [32]–[34]. In [21], it has been numerically demonstrated that a purely reactive load, *i.e.* an inductor or a capacitor, that is mismatched with an ideal feeding transmission line, can be transformed into a resistance $R$ when a proper temporal modulation profile is applied to the inductance $L(t)$, or capacitance $C(t)$, of the load (Fig.1(b)).

If the temporal modulation is present, the energy is accumulated within the reactive load without generating any reflection towards the source, leading to a perfect matching. Such an approach requires extremely fast modulation of the reactive components and, therefore, the implementation complexity prevents its realization in realistic scenarios, despite some interesting attempts have been provided [35].

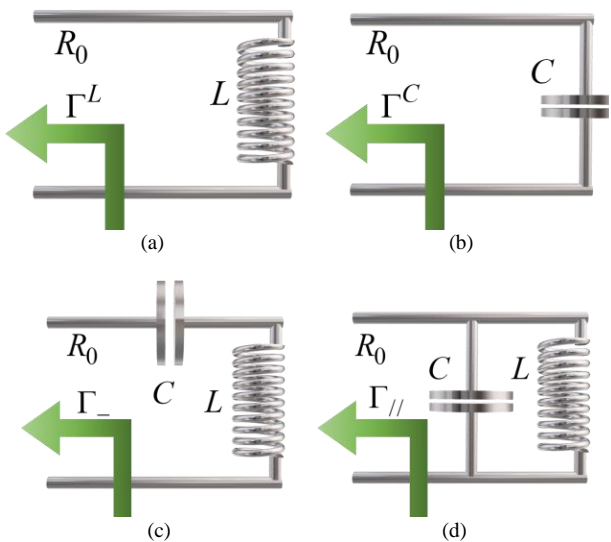

Fig. 2. Schematics of a resistive transmission line with characteristic impedance $R_0$ loaded with a single reactive element, an inductor $L$ in (a) and a capacitor $C$ in (b) and a connection of $LC$ elements in (c) series and (d) parallel.

In this work, we reply to the following intriguing question: *can the temporal profile of the propagating signal enable a special impedance transformation, such that impedance matching of a purely reactive load terminating an ideal transmission line, as shown in Fig.1(c), is achieved*? A decade ago, the use of temporal profile of incoming signal has been exploited in [36]–[38] for enabling efficient coupling in single atoms and optical resonators. More recently, the same concept has been moved in optics, for enabling energy virtual absorption in lossless systems, making them appear as matched to free-space if complex zeros of its scattering eigenmodes are excited [39]–[45], reporting also elastodynamic experimental proof in mechanical systems [46]. In particular, the authors have investigated in [45] the possibility to enable virtual absorption effects in metasurface-bounded cavities. The analysis of this problem led to the derivation of the complex zeros, and corresponding limit values of the surface impedances of the metasurface. Even though the concepts of virtual absorption [39] and virtual critical coupling [44] have already been investigated in cavities and optical applications, here, we exploit it for achieving a similarly anomalous response also in microwave circuits composed by purely reactive lossless lumped elements, leading to a *virtual perfect matching*, and opening the door to a new impedance matching strategy, as preliminarily shown in our recent conference work [47], [48].

This contribution aims at shedding light for the first time on this phenomenon from both the physical and the electrical point of view, through a detailed theoretical analysis, numerical and experimental results, deriving the fundamental limits of the impedance matching concept based on time-modulated excitation signals. We analyse all possible configurations of reactive loads reported in Fig. 2: single inductance (Fig.2(a)) and capacitance (Fig.2(b)), and series/parallel connection of those elements as reported in Fig.2(c)-(d), respectively. The parallel connection of reactive loads has been already discussed in our recent work [45], where fundamental limits of virtual perfect absorption have been analysed. Nevertheless, this study included preliminary analytic expressions, and it was related to free-space propagation, focusing on energy accumulation of the structure.

This paper reveals that purely reactive loads can produce no reflection when connected to a lossless transmission line exhibiting a resistive characteristic impedance under special excitation conditions. The specific novelties introduced by this work in the state of the art can be summarized in the following four contributions:

*Contribution 1)* We develop and discuss in detail a general theoretical analysis of all element configurations available as reactive loads, deriving the operative regions for which the phenomenon takes place as a function of the natural frequency of the circuit and the circuit time constant.

*Contribution 2)* We provide a set of proper numerical simulations of all presented load cases, describing the physical insights and providing a comparison among them.

*Contribution 3)* We demonstrate that the analyzed phenomenon is not related to a delay along the line, setting a numerical simulation based on an electrically long microstrip line.

*Contribution 4)* We report the experimental results of perfect matching condition for the series and parallel configurations.

According to the aforementioned list, the paper is organized as follows: in sections II the theoretical analysis of the mismatching condition under complex excitation is reported and discussed (*Contribution 1*), considering all the different load configurations (single reactive loads in Section II.A and series and parallel connections in Section II.B). Section III is devoted to the theoretical results discussion and to the constraints and limits evaluation on the load values for achieving the virtual perfect matching (*Contribution 2*). In Section IV, we report the numerical simulations for purely imaginary and complex excitations, and the demonstration that the phenomenon is not related to a delay along the line (*Contribution 2 and 3*). Finally, Section V presents the experimental results, demonstrating that the phenomenon can be observed in lumped microwave networks (*Contribution 4*). In Section VI we draw the conclusions of this investigation.

## II. REACTIVE LOAD UNDER TIME MODULATED EXCITATION: CIRCUITAL ANALYSIS

Let us consider a circuit where a lossless transmission line with resistive characteristic impedance $R_0$ is loaded by a generic ~~reactive~~ network. From basic transmission line theory, we can write reflection coefficient seen at the load as:

$$\Gamma(\omega) = \frac{Z_L(\omega) - R_0}{Z_L(\omega) + R_0}. \tag{1}$$

Regardless of the complexity of the network, the load impedance can be always represented as $Z_L(\omega) = R_L(\omega) + jX_L(\omega)$, where $X_L$ is an effective frequency-dependent reactance, seen at the load terminals. Considering a purely reactive load network, thus having $\text{Re}[Z_L] = R_L = 0$ in eq. (1), we can write the reflection coefficient amplitude as:

$$|\Gamma(\omega)| = \left|\frac{jX_L(\omega) - R_0}{jX_L(\omega) + R_0}\right|. \quad (2)$$

In general, this lossless transmission line is always mismatched to the load, since $X_L$ and $R_0$ will always have different values in the real frequency spectrum. Thus, the imaginary quantity $jX_L(\omega)$ should be converted into a real one and equated to the real characteristic impedance $R_0$ of the feeding transmission line. Exploiting the temporal modulation of the incoming signal, this can be done by introducing a complex value of the excitation frequency $\omega$, i.e. $\omega = \omega_r + j\omega_i$ [41]. A complex frequency excitation is a harmonic signal of frequency $\omega_r$, whose amplitude temporally varies following an exponential growing (or decaying) profile proportional to $e^{\omega_i t}$.

The reflection coefficient in eq. (2) can be represented as a function of the frequency in the complex frequency plane, where it is possible to identify scattering singularities, i.e. zeros and poles, of the amplitude of the reflection coefficient function for specific combinations of real and imaginary frequency. Hereafter, we state this analysis for a single reactance case and a combination of LC in series and parallel connection loads. Inductors and capacitors are linear components, in accordance with Ohm's law even when the applied voltage signals increase due to the exponential growing in amplitude dictated by the complex frequency. Non-linearity in the circuit components may represent a further degree of freedoms for these phenomena, not considered in this work.

*A. Single load cases*

Now, we consider a load consisting of a single inductor $L$ or capacitor $C$, whose impedance is $Z_L = j\omega L$ or $Z_L = 1/j\omega C$, respectively (Fig.2(a)-(b)). Starting from eq. (2), we can estimate the reflection coefficients at the load terminals for the two considered cases:

$$\Gamma^L(\omega) = \frac{j\omega\tau - 1}{j\omega\tau + 1} \quad (3)$$

$$\Gamma^C(\omega) = -\Gamma^L(\omega), \quad (4)$$

where superscripts "$L$" and "$C$" identify the cases reported in Fig.2(a)-(b), respectively, and the quantity $\tau$ assumes the values $\tau = L/R_0 = \tau^L$ in eq. (3) and $\tau = R_0 C = \tau^C$ in eq. (4). In circuit theory, the quantities $\tau^L$ and $\tau^C$ are the well-known circuit time constants for a transient step response for $RL$ and $RC$ circuits, and their value determines how fast the time response goes to zero.

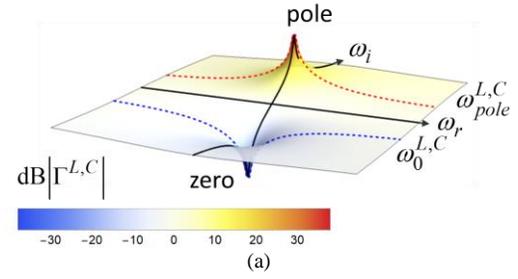
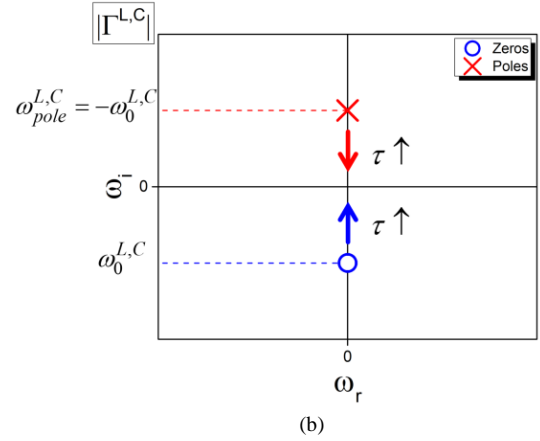

Fig. 3. Reflection coefficient amplitude in the complex frequency plane for a single reactive load, inductive or capacitive. In (a), its 3D representation in dB showing diverging singularities on the imaginary frequency axis. In (b), singularities position and behavior according to time constants.

We now consider the complex frequency excitation of the two circuits in Fig.2(a)-(b) by imposing $\omega = \omega_r + j\omega_i$ in eqs. (3)-(4). Fig.3(a) reports the amplitude of the reflection coefficient in the complex frequency plane, that exhibits two singularities: a pole (Fig.3(a), red/yellow positive peak) is in the positive half-space of the imaginary frequency, whereas a zero (Fig.3(a), blue negative peak) is in its negative half-space. The reflection coefficient pole excitation leads to an anomalous condition for which the reflected energy is higher than the impinging one. In passive systems, this is as a special scattering condition for which the scattered field decays slower than the excitation field, giving rise to a virtual gain effect [49]. On the other side, exciting the zero implies a zero-reflection condition: the load behaves as an indefinite accumulator for the illuminating signal, without dissipating its energy but rather storing it within itself. The zero-scattering condition is satisfied as long as the complex frequency is such to engage the scattering zero, thus leading to a zero-reflection coefficient. Modifying the signal excitation frequency, the reflection coefficient changes and assumes a non-zero value, energy then leaks out from the reactive load.

Imposing eqs. (3)-(4) equal to zero, we derive the roots of the reflection coefficients $\Gamma^{L,C}$ with respect to frequency:

$$\omega_0^{L,C} = \omega_r + j\omega_i = 0 - j\tau^{-1}, \text{ with } \tau = \tau^{L,C} \quad (5)$$

The reflection coefficient vanishes at purely imaginary frequencies $\omega_0^{L,C}$, as expected from Fig.3(a). An excitation signal with an imaginary frequency is not oscillating, but simply growing over time with a steepness given by the exponential factor $e^{-\omega_0^{L,C}}$. This is fully consistent with circuit theory where the charging and discharging behavior of a *RC* (*RL*) circuit is described by the same exponential curve with a factor $e^{-t/\tau}$ for voltage (current). It is clear, therefore, that the circuit time constant $\tau$ plays a fundamental role in the zero and pole positions in the complex frequency plane. In a reciprocal lossless system, zeros and poles are always complex conjugate; increasing the time constant $\tau$, both zero and pole move inside the complex frequency spectrum, reducing the required steepness of the time-growing signal for being excited (Fig.3(b)), always conserving complex conjugation.

### B. Series and parallel load cases

Let us now consider the load as a reactive network modelled as a series/parallel connection of an inductor and capacitor (Fig.2(c)-(d)). In these two scenarios, the load reactance can be found as:

$$jX_L^- = j\frac{\omega^2 - \omega_{res}^2}{\omega C \omega_{res}^2}; \qquad jX_L^{//} = j\frac{\omega L \omega_{res}^2}{\omega_{res}^2 - \omega^2} \qquad (6)$$

where the superscripts "-" and "//" identify the series and parallel connection, respectively, and $\omega_{res} = 1/\sqrt{LC}$ is the natural resonant frequency of the *LC* circuit for both configurations. Substituting eq. (6) into eq. (2), we obtain:

$$\Gamma^-(\omega,\tau^-) = \frac{(\omega^2 - \omega_{res}^2) + j(\tau^- \omega_{res}^2)\omega}{(\omega^2 - \omega_{res}^2) - j(\tau^- \omega_{res}^2)\omega} \qquad (7)$$

$$\Gamma^{//}(\omega,\tau^{//}) = -\Gamma^-(\omega,\tau^-) \qquad (8)$$

where $\tau^- = R_0 C$ ($\tau^{//} = L/R_0$) are for series (parallel) configuration. By forcing eqs. (7)-(8) to vanish, we obtain the following perfect matched complex frequencies:

$$\omega_0^{-,//} = -\frac{j}{2}\omega_{res}\left(\tau\omega_{res} \pm \sqrt{\tau^2 \omega_{res}^2 - 4}\right), \quad \text{with} \ \tau = \tau^{-,//} \quad (9)$$

The frequencies granting zero-reflection are two due to "$\pm$" sign in eq. (9) and seem to be purely imaginary, as in the case of a single reactive load, however, this is true only if the argument of the square root in eq. (9) is positive, *i.e.*, $\tau^2 \omega_{res}^2 - 4 \geq 0$. When this condition is not fulfilled, that is $\tau^2 \omega_{res}^2 - 4 < 0$, eq. (9) assumes a complex value, that allows achieving a complex frequency with non-zero real part, *i.e.* harmonic signals at frequency $\omega_r$ with temporal varying

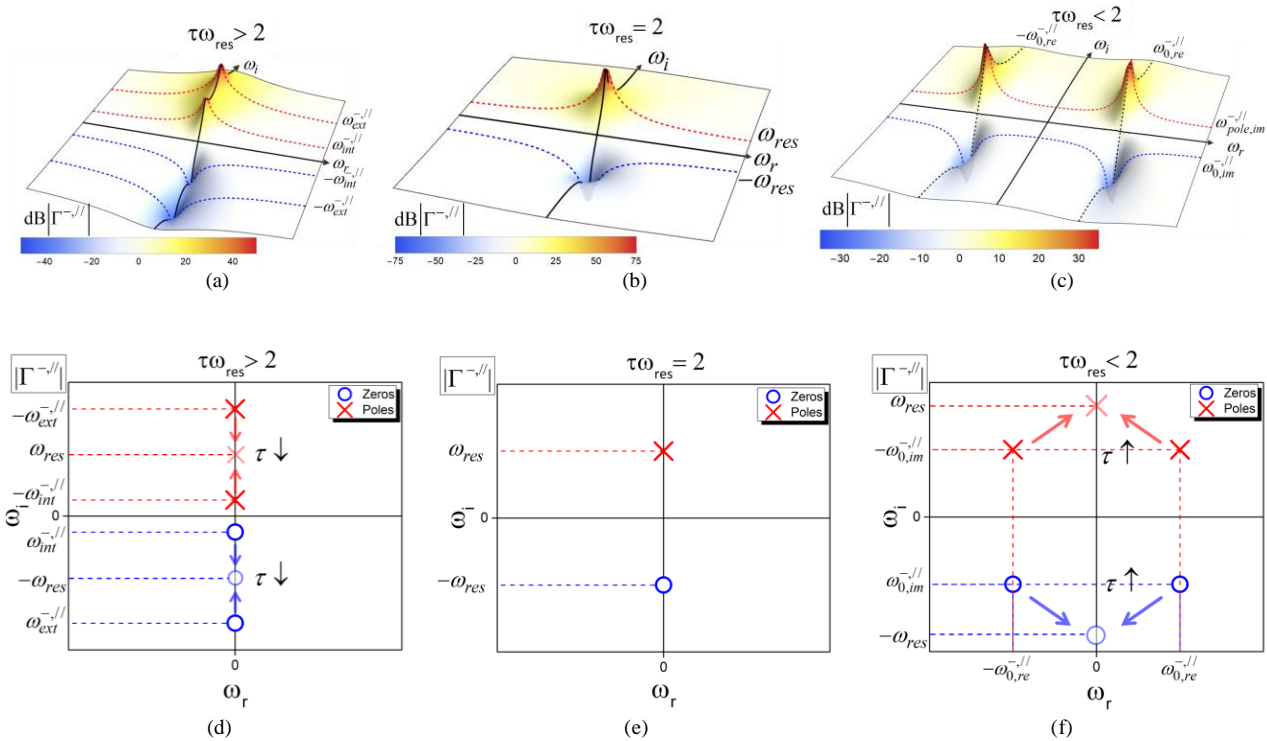

Fig. 4. 3D representation of the reflection coefficient amplitude for series/parallel *LC* loads in the complex plane (a)-(c) show zeros in lower half-plane and poles in higher half-plane according to time constants. (d)-(f) show quality behavior of singularities, which are: (d) purely imaginary when $\tau\omega_{res} > 2$, (e) coincident purely imaginary when $\tau\omega_{res} = 2$, (f) complex when $\tau\omega_{res} < 2$.

envelope, growing or decaying according to the sign of $\omega_i$.

Let us now plot and discuss the reflection coefficient complex frequency planes for the following cases, being the quantity $\tau\omega_{res}$ always positive: 1) $\tau\omega_{res} > 2$, 2) $\tau\omega_{res} = 2$ and 3) $\tau\omega_{res} < 2$. In all three cases, the reflection coefficient always shows singularities, but their location significantly changes according to the current value of the product $\tau\omega_{res}$.

Case 1) consists of a LC circuit whose $\tau\omega_{res} > 2$. In this scenario, the system exhibits two separate frequencies for which zero-reflection can be achieved (Fig.4(a)-(d)). The reflection coefficient has two zero-pole pairs symmetrically distributed along the imaginary frequency axis at $\pm\omega_{int}^{-,//}$ and $\pm\omega_{ext}^{-,//}$, where the subscripts *int* (*ext*) identify the zeros that are closer (farther) from the real frequency axis. Like single load case, the system exhibits zero-reflection only if the excitation signal is not oscillating. However, in this case, the loci of the zeros and poles of the reflection coefficient move differently with respect to the value of the circuit time constant $\tau$. For higher values of the time constant, the external zero-pole pair at $\pm\omega_{ext}^{-,//}$ moves farther, whilst the internal one at $\pm\omega_{int}^{-,//}$ tends to the frequency plane origin. On the contrary, when $\tau$ decreases (Fig.4(d)), both external and internal zero-pole pair move towards the imaginary counterpart of the natural frequency, $\pm j\omega_{res}$, which involves special responses from the system as discussed in next cases.

In case 2), when $\tau\omega_{res} = 2$, the two zero-pole pairs of previous case degenerate to the imaginary counterpart of the natural resonant frequency $\pm j\omega_{res}$ (Fig.4(b)). The degeneration of the two zero-pole pair into two imaginary frequencies $\pm j\omega_{res}$ is confirmed in eq. (9), where $\omega_0^{-,//} = -j\omega_{res}$ for $\tau\omega_{res} = 2$, and, consequently, $\omega_{pole}^{-,//} = +j\omega_{res}$ (Fig.4(d)). Under this condition, then, the natural resonant frequency value $\omega_{res}$ is not used to express the circuit oscillation as predicted in circuit theory, instead, it coincides with the exponential factor to provide perfect matching, thus $e^{-\omega_i t} = e^{t/\sqrt{LC}}$.

The last case consists of a LC circuit whose $\tau\omega_{res} < 2$. In Fig.4(c), the reflection coefficient exhibits two zero-pole pairs for complex frequencies with a non-zero real part, demonstrating that it is possible to achieve zero-reflection from a purely reactive load under monochromatic excitation at $\omega_r$, provided that the amplitude of the wave follows the profile $e^{-\omega_i t}$. It is worth noticing that Fig.4(c) reports two possible zeros symmetrically distributed with respect to the imaginary axis, but only the zero for positive real frequency can be excited. Hence the left zero-pole pair can be neglected hereafter. The singularities behavior, in this case, is showed in Fig.4(f): keeping the same $\omega_{res}$ and increasing $\tau$, the zero-pole pair tends to the two coincident values $\pm j\omega_{res}$, leading to case $\tau\omega_{res} = 2$.

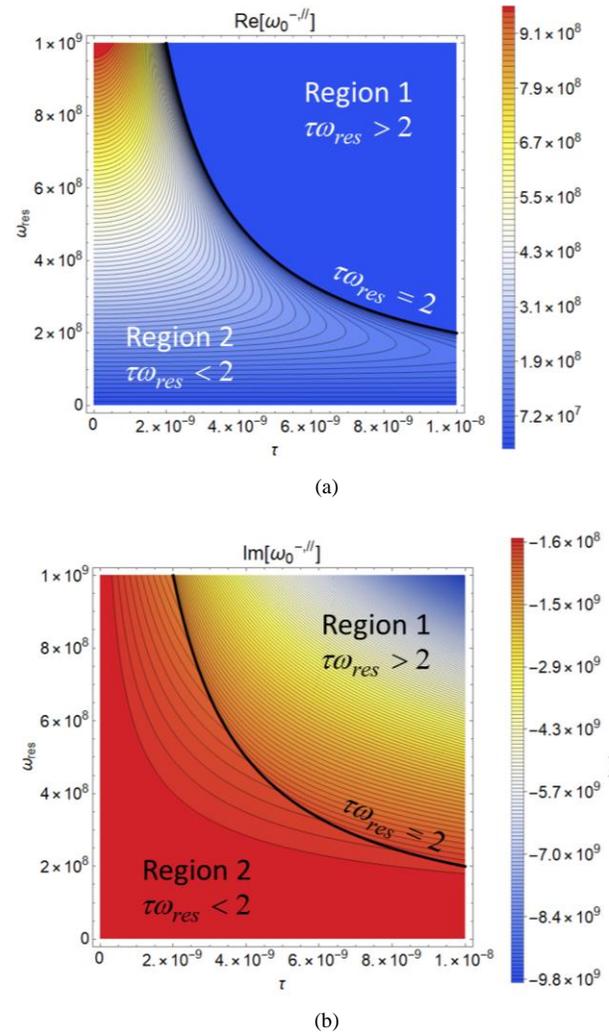

(a)

(b)

Fig. 5. In LC load configurations, the real (a) and imaginary (b) parts of the scattering zeros frequency with respect to time constant and natural resonance frequency are showed in contour plots. The black line corresponds to $\tau\omega_{res} = 2$ and splits the scattering zeros plots in Region 1 where $\tau\omega_{res} > 2$ and Region 2 where $\tau\omega_{res} < 2$.

### III. DISCUSSION OF RESULTS AND PHYSICAL INSIGHTS

From this analysis, we can state that the imaginary resonant frequency $j\omega_{res}$ is a key parameter, since it is the boundary frequency value between the two regions, and it is strongly related to the natural real resonant frequency of the LC circuit. From the analytical point of view, at $\omega_0 = -j\omega_{res}$, the reactive load terminating the transmission line exhibits the same real value of the transmission line characteristic impedance, i.e., $Z^{-,//}(\omega_0) = jX^{-,//}(-j\omega_{res}) = R_0$. Imposing $\omega_0 = -j\omega_{res}$ into eqs. (6)-(7) and the reflection coefficient equal to zero, we derive the values shown by the load at any instant of time as long as the signal is applied, i.e. $Z^- = 2\sqrt{L/C} = R_0$ and $Z^{//}=0.5\sqrt{L/C}=R_0$ for the series and parallel case, respectively. From the physical point of view, all the energy provided to the load under exponentially growing signal is stored inside reactive components (with opposite sign), which are both gradually charged in time. The energy will be accumulated progressively, and, at the same time, the voltage-current ratio at the load terminals always equals the transmission line resistance, thus forbidding reflections back to the source.

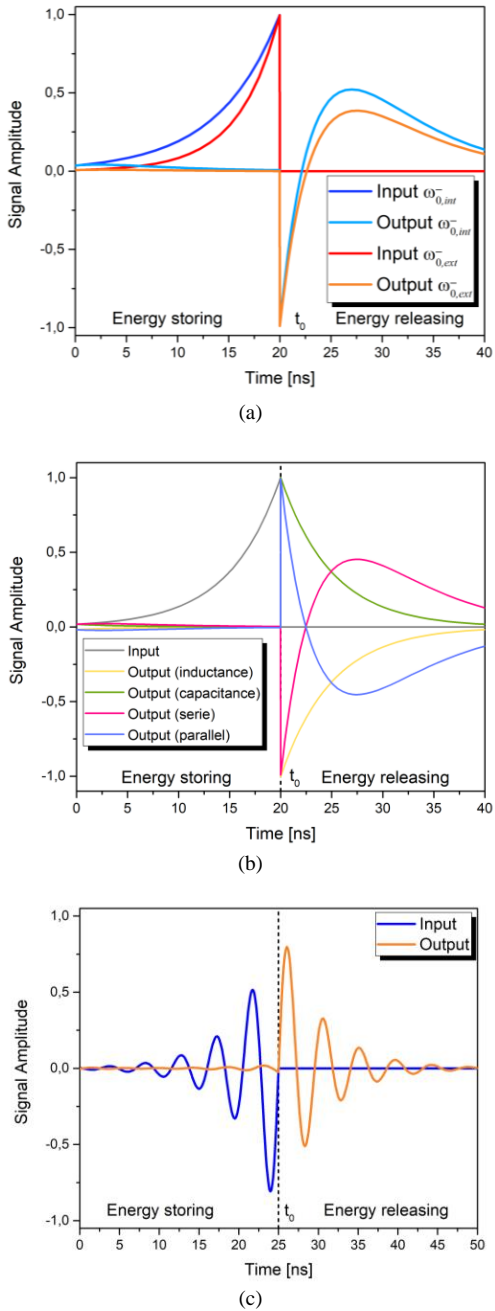

Fig. 6. Transient response for: (a) $LC$ series case with $L=120\text{nH}$ and $C=200\text{pF}$ for $\tau\omega_{res}>2$ case (two imaginary zeros), (b) four different loads satisfying $\tau\omega_{res}=2$ whose parameters values are in Table I, (c) $LC$ parallel load with $L=10\text{nH}$ and $C=50\text{pF}$ for $\tau\omega_{res}<2$ case.

To provide further validation to our theoretical investigation, we can see that the proposed theoretical analysis resembles the damping behavior of $RLC$ circuits, where the circuit behaves as overdamped, critically damped or under damped according to the roots (complex or real) of the circuit harmonic oscillator equation [50], [51]. In circuit theory, damping is caused by the energy dissipation inside the resistance of an $RLC$ network. We stress that, in our study, the resistance of the transmission line $R_0$ does not dissipate energy leading to circuit damped response. Here, instead, we exploit complex excitation (having the role of resistance dissipation) to balance the corresponding circuit reflections, that are zero in the transient response.

We now illustrate the behavior of the real and imaginary part of zeros for both $LC$ series and parallel connections, thus $\omega_0^{-,//}$ as expressed in eq. (9), with respect to the two parameters $\tau$ and $\omega_{res}$. Fig.5(a)-(b) show that both real and imaginary parts of $\omega_0^{-,//}$ plots are split in two regions, delimited by the hyperbolic curve $\tau\omega_{res}=2$. As expected, the real part of $\omega_0^{-,//}$ (Fig.5(a)), is zero when $\tau\omega_{res}\geq 2$, (Region 1) meaning that zeros are purely imaginary, as predicted by eq. (9). In the same Region 1, the imaginary part in Fig.5(b) exhibits very high frequency values, meaning that a sharper steepness is needed to engage the desired scattering zero. This region is not of interest for seeking virtual absorption with modulated signals. On the contrary, when $\tau\omega_{res}<2$ (Region 2), the real part is non-zero and, for a given $\tau$, it is directly proportional to the $LC$ resonant frequency $\omega_{res}$. The imaginary part is, indeed, much smaller than the real part, meaning that the steepness is reduced, leading to an exponential slowly growing signal.

IV. SIMULATIONS RESULTS

The impedance matching achieved through the anomalous transient behavior of reactive loads under complex excitation has been numerically verified through CST Studio Suite [52]. We consider a lossless 50-Ohm transmission line with different loads terminating it. In circuit schematics, the impedance transmission line coincides with the generator internal resistance. In next sub-sections, we will focus on the circuital schematic simulations of single and combined loads that show perfect matching under purely imaginary and complex frequency excitation cases in Sections IV-A and IV-B, respectively. Finally, in Section IV-C, full-wave simulations over a designed microstrip line are exposed.

A. Imaginary excitation cases

In $\tau\omega_{res}>2$ case, there are two purely imaginary scattering zeros that can be engaged to get perfect matching. We choose a series load composed by $L=120\text{nH}$ and $C=200\text{pF}$, satisfying $\omega_{0,int}^{-}=-j1.67\times 10^8\text{rad/s}$ and $\omega_{0,ext}^{-}=-j2.5\times 10^8\text{rad/s}$ zeros, whose input signals are in Fig.6(a), red and dark blue lines. The incident and reflected signals are recorded at the load terminals. As expected, the impinging signals are engaging the reflection coefficient zeros and no reflection takes place. Despite the entire circuit is passive, energy cannot be dissipated since the load is purely reactive, allowing only energy storing in the reactive load. As soon as the signal stops, at kick-off time $t_0$, the zero-reflection condition is not satisfied anymore, forcing the reactive load to release the stored energy (Fig.6(a), orange and light blue lines). We can furthermore observe that both zeros satisfy the circuit, and the zeros present different growing envelope according to the $\omega_0^-$ values.

To explore single reactance and critical ($\tau\omega_{res}=2$ for series/parallel connection) cases, we choose inductors and capacitors values as described in Table I, all of them satisfying the same zero $\omega_0 = -j2\times10^8$ rad/s. The signal is not oscillating in time as predicted by eqs.(5)-(9), for $\tau\omega_{res}=2$ (Fig.6(b), grey line). The circuits transient response verifies perfect matching condition.

Reflected output signals are zero until kick-off instant $t_0$, but after it the accumulated energy is released. Each output shows a different behavior according to the circuit to which it is referred to, and corresponds to *RC*, *RL*, *RLC* series/parallel circuits where the charging and discharging behavior is described by the same exponential curve with a factor $e^{-t/\tau}$ for voltage and current, respectively.

TABLE I
LOAD VALUES UNDER $e^{j\omega_0 t}$ EXCITATION
($\omega_0 = -j2\times10^8$ rad/s)

| Load type | Inductance [nH] | Capacitance [pF] |
|---|---|---|
| L | 250 | --- |
| C | --- | 100 |
| LC series | 125 | 200 |
| LC parallel | 500 | 50 |

*B. Complex excitation cases*

To verify $\tau\omega_{res}<2$ case, we choose an *LC* parallel configuration where $L=10\text{nH}$ and $C=50\text{pF}$. According to eq.(9), the reflection coefficient zero is $\omega_0^{//} = (1.4\times10^9 - j2\times10^8)$ rad/s, *i.e.*, a monochromatic signal at frequency $f_r^{//} = 222\text{MHz}$ with a time-varying growing envelop following the profile $e^{2\times10^8 t}$ (Fig.6(c), blue signal) Perfect matching is achieved also for complex valued frequencies, showing zero reflections up to kick-off instant (Fig.6(c), orange line).

*C. Full-wave microstrip verification*

To give further validate the observed phenomenon of perfect matching under complex excitation, we have considered a realistic microstrip line printed on a 0.5mm-thick FR4 substrate ($\varepsilon_r=4.3$) and terminated on the parallel *LC* load. The microstrip line depicted in Fig. 7(a) has been designed at $f_0 = 5GHz$. The microstrip is made of copper with width $d=0.9mm$. The goal of this numerical analysis is to demonstrate that this phenomenon is *not related to a propagation delay along the line*, but to energy accumulation and release within the reactive load. The microstrip line has been chosen of a reasonable length $l \approx 66mm$ to let the reader appreciate the whole excitation on the line and the accumulation of energy. The loaded elements are a parallel *LC* where $L=0.5nH$ and $C=2pF$. Referring to eq. (9), a complex zero is required, working at the microstrip operative real frequency $f_0$. The exponential envelope needed to excite the reflection coefficient complex zero corresponds to the imaginary frequency $\omega_i = -5\times10^9$ rad/s.

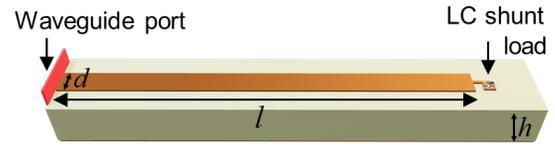

(a)

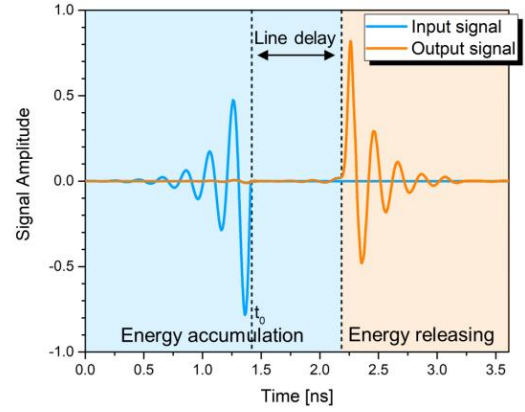

(b)

Fig.7. (a) Schematic of the microstrip line system with an *LC* shunt load with $L=0.5nH$ and $C=2pF$ excited at waveguide port by a time-modulated excitation $\omega_0 = 2\pi f_0 - 5\times10^9$. (b) Incident and released signals recorded at the port.

In Fig.7(b) we report the recorded signals at the port terminals over time. The blue curve represents the incident signal, and the orange curve the released signal. It is clear that the phenomenon is preserved, since 1) the envelope of the released signal is opposite with respect to the incident one, whereas in case of reflection it would not, and 2) the presence of the time delay between the two signals is due to their propagation along the line. Since the signals are detected at waveguide port, the observed delay exactly corresponds to two times the line length over the phase velocity in microstrip line, that is:

$$\Delta t = 2l \frac{\sqrt{\varepsilon_{eff}}}{c} \qquad (10)$$

where $\varepsilon_{eff}$ is the effective permittivity of the microstrip line and $c$ is light speed. The line delay naturally occurs during propagation over the microstrip, and it is independent from the perfect matching phenomenon. These numerical results clearly show that the signal is not only delayed because of the line length, but it is independently stored thanks to perfect matching due to the excitation of the reflection coefficient zeros in the complex frequency plane.

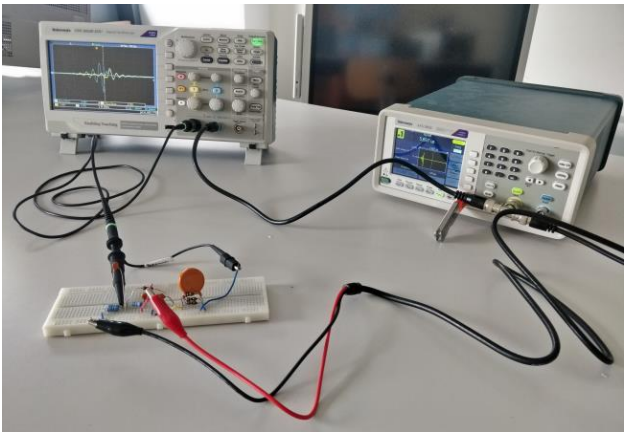

Fig. 8. Experimental setup with Tektronix TBS1052EDU oscilloscope (on the left), breadboard with components (centre) and Arbitrary Function Generator Tektronix 1062 (60MHz) (on the right).

## V. EXPERIMENTAL VERIFICATION

In the following, we set up an experiment for validating our analysis and demonstrate the perfect matching by using a reactive load consisting of an $LC$ parallel and an $LC$ series connection, as detailed in subsections V.A and V.B, respectively. In both configurations we use the same components: a ceramic disc capacitor of $C_{eq} = 2nF$ and three axial lead inductors of $L = 1\mu H$ that show, in shunt connection, an equivalent of $L_{eq} = 333nH$. The use of multiple inductors or capacitors that correspond, in shunt or series connections, to an equivalent value, does not affect the phenomenon, that is preserved. A picture of the experimental set is shown in Fig. 8, where the chosen components are mounted over a breadboard, connected to the arbitrary function generator (AFG) Tektronix 1062 that generates the time modulated signals. On the other side, an oscilloscope Tektronix TBS1052EDU is connected to a probe to the board and detects the voltage signals.

### A. Parallel LC load

Following eq. (8), the shunt connection of $L_{eq}$ and $C_{eq}$ falls under condition $\tau^{//}\omega_{res} < 2$, so two complex zeros arise on the reflection coefficient complex plane as shown in Fig.4(c)-(f). The chosen complex frequency zero needed to get perfect matching has $f_r = 6.1 MHz$ with a temporal modulation composed of an exponential rise given by the imaginary frequency $\omega_i = -5\times 10^6$ rad/s.

As shown in Fig. 9a, the load is connected through a lumped-element Wilkinson Power Divider (WPD) designed at $f_r$ to the AFG and to the oscilloscope. The signal is generated by the AFG and is split in two signals, the former, CH1, is directly connected to the oscilloscope, and CH2 is connected to the WPD and combined into the line leading to the load. Here, the signal is accumulated and, later, is divided into the two ports that carry the signal back to the AFG and to the oscilloscope. Both the output impedance of AFG and oscilloscope are set at $R_0 = 50\Omega$, also the oscilloscope probe lead measures voltage over a $50\Omega$ load. In Fig.9(b)-(c), the simulated response of the circuit detected at the probe port and the measured response detected at AFG port are reported. As expected, the experiment clearly returns perfect matching phenomenon since the signal detected at the probe is zero up to kick-off instant $t_0$, that is, reflections vanish. After $t_0$ instant, as soon as the temporal modulated excitation stops, the signal is released and is equally split by the power divider into CH1 and CH2 signals, being the perfect matching condition not satisfied anymore. A negligible difference between simulated and measured released signals can be observed comparing the yellow curves in Fig. 9(b)-(c), due to the very small resistive behavior introduced by wires and connectors in the experiment that leads to a minor attenuation of the signals.

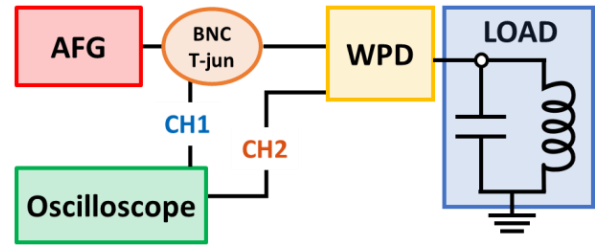

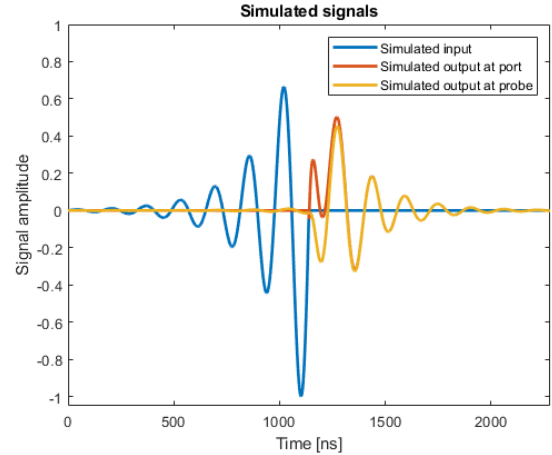

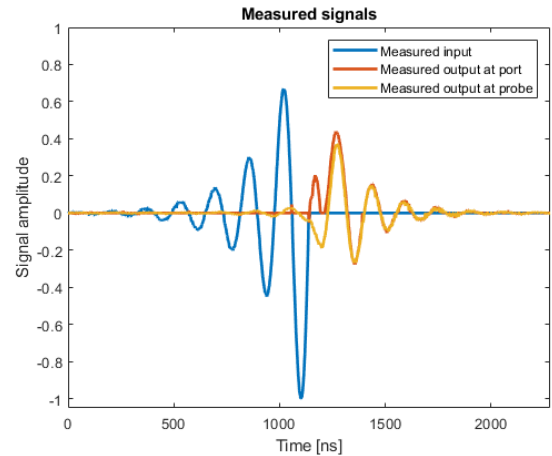

Fig. 9. (a) Schematic of the system for LC shunt load. Normalized simulated (b) and combined measured (c) input signal (blue curve), output signal at port (CH1) reflected towards the AFG port (red curve) and output signal detected at the oscilloscope probe (CH2) (red curve).

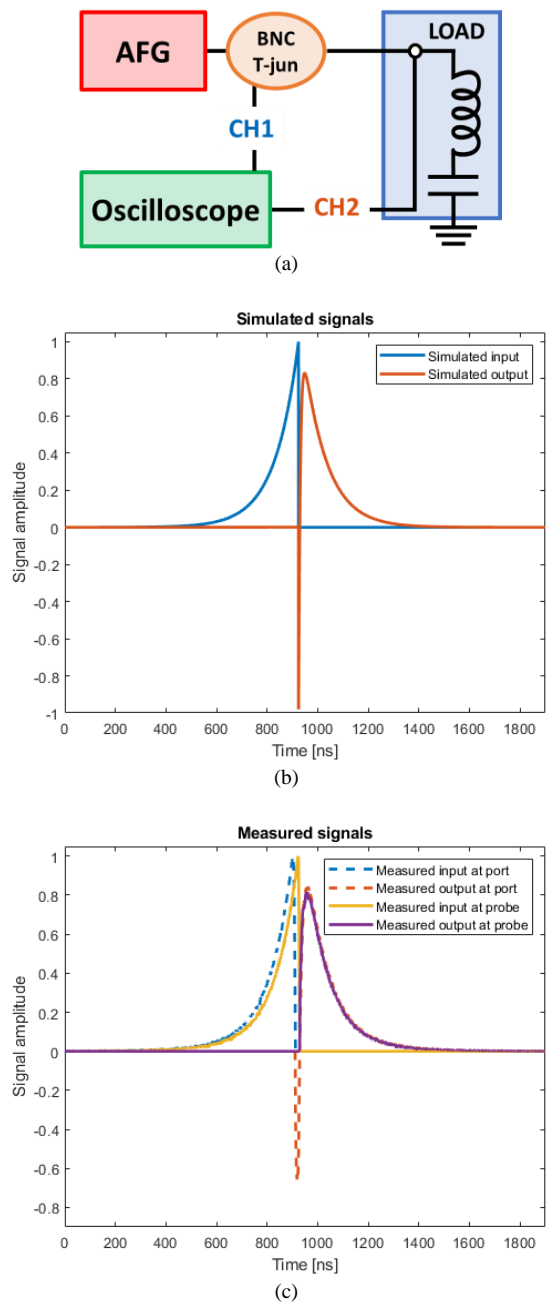

Fig. 10. (a) Schematic of the system for LC series load. (b) Normalized simulated input and output signals. (c) Normalized combined measured signals showed at port (CH1): input signal (blue dashed curve) and output signal (red dashed curve). Combined measured signals showed at probe (CH2): input signal (yellow curve), reflected towards the AFG port (red curve) and output signal detected at the oscilloscope probe (CH2) (purple curve).

### B. Series load

In addition to the shunt *LC* load examination, a second analysis for a *LC* series connection has been carried. Here, the considered load is composed by a series connection of $L_{eq}$ and $C_{eq}$ elements. The series configuration of these two elements provides $\tau^- \omega_{res} > 2$, in eq. (9). Consequently, perfect matching is obtained through two purely imaginary zeros, that are $\omega_{int} = -j1.08 \times 10^7$ rad/s and $\omega_{ext} = -j1.41 \times 10^8$ rad/s, using the same notation of Section II. These zeros lay along the imaginary frequency axis as shown in Fig.4(a)-(d). We have arbitrarily chosen $\omega_{int}$ zero to excite the circuit. Since now the excitation $\omega_{int}$ required to get perfect matching is a static growing signal, the WPD element is not needed anymore, as shown in the experimental setup in Fig.10(a). As for the shunt connection case, CH1 carries the port signal, that comes from the BNC splitter and directly connects to the oscilloscope. CH2 carries the probe signal, detected by the probe on the breadboard. In this case, port and probe signals coincide in simulations, since retards and small dissipation are not taken into account. However, since some delays and dissipations in measures naturally occur, we show both measured port and probe signals. In Fig. 10(b) and (c), we respectively show simulated normalized signals and the combined measured signals. As for series case presented in Fig.6(b) (pink curve), perfect matching behaves similarly, in both simulated and measured results. As for the measured signals in Fig. 10(c), signals at the probe exhibit an extremely small delay with respect to the ones at the port. However, it is enough to hide the negative peak at the probe that is indeed present in the simulated signals. The accuracy and sampling speed of measuring instruments have been optimally set for detecting any variation of the signal amplitude at the measuring points. Indeed, the extremely narrow negative peak curve, presented by the output signal at the port (orange dashed signal in Fig.10(c)), is correctly detected. The absence of the negative peak on the measured signal at the probe is, thus, due to the intrinsic charging and discharging time required by reactive circuital components, that in this case is longer than the duration of such a peak.

## VI. CONCLUSION

To conclude, we have presented the first experimental validation and full theoretical derivation of perfect matching of purely reactive loads terminating a lossless transmission line by exploiting the concept of impedance transformation through time-varying complex signals. In particular, the work sheds light on the anomalous phenomenon of virtual perfect matching, bringing several novelties to the current state of the art of complex signals propagation in electromagnetic systems modelled through terminated transmission lines, and, for the first time, proposing an experimental implementation. We have developed and discussed in detail a general theoretical analysis of all element configurations available as reactive loads, deriving the operative regions for which the phenomenon takes place as a function of the circuit natural frequency and time constant. The theoretical analysis has been verified through a set of proper numerical simulations of all presented load cases, describing all physical insights and providing a comparison among them. Moreover, we have clarified, by setting a numerical simulation based on an electrically long microstrip line, that the analyzed phenomenon is not related to the delay along the line. Finally, we have reported the first experimental results on virtual effects and, in particular, perfect matching condition for the series and parallel configurations. Although the required increasing envelope of the temporal profile is still a limit for making this concept exploitable in practical scenarios, the experimental verification provided in this paper may provide a base for paving the way to a new impedance matching strategy in microwave circuits, that can be exploited together with the already proposed applications based on virtual effects such as wireless energy transfer and energy harvesting [53]–[55].

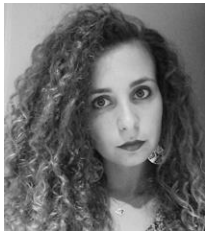

**ANGELICA VIOLA MARINI** (S'17) was born in Rome in 1992. She received the M.M. degree in Piano Performance from Conservatorio di Musica Santa Cecilia, Rome, Italy and the B.S. and M.S. degrees in electronics and ICT engineering from ROMA TRE University, Rome, Italy, in 2012, 2014 and 2017, respectively. She is currently pursuing the Ph.D. degree in the Department of Applied Electronics at ROMA TRE University, Rome, Italy.

She made an internship as Engineer in Wind Tre S.p.A. for 5G deployment experimentation in 2017. Her research fields are time-modulated phenomena and structures, and anomalous electromagnetic propagation enabled by metasurfaces and microwave components.

Dr. Marini is a member of the Virtual Institute for Artificial Electromagnetic Materials and Metamaterials (METAMORPHOSE VI, the International Metamaterials Society). She served as a Member of local organizing committee of Congress Metamaterials'2019, held in Rome, Italy, in 2019 and as Member of local organizing committee of XXXVIII and XXXIX Euprometa Doctoral schools, held in Rome, Italy in 2017 and 2019, respectively.

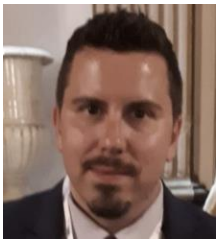

**DAVIDE RAMACCIA** (S'11–M'14–SM'19) received the B.S. (summa cum laude) and M.S. (summa cum laude) degrees in electronic and ICT engineering and the Ph.D. degree in electronic engineering from ROMA TRE University, Rome, Italy, in 2007, 2009, and 2013, respectively. Since 2013, he has been with the Department of Engineering (2013-2021) and then, with the Department of Industrial, Electronic, and Mechanical Engineering (2021-now), at ROMA TRE University. His main research interests are in the modelling and design of (space-)time-varying metamaterials and metasurfaces, and their applications to microwave components and antennas, and the analysis of anomalous scattering effects in temporal metamaterials. He has coauthored more than 100 articles in international journals, conference proceedings, book chapters, and holds one patent.

Davide Ramaccia has been serving the scientific community, by playing roles in the management of scientific societies, in the editorial board of international journals, and in the organization of conferences and courses. He is currently a General Secretary of the *Virtual Institute for Artificial Electromagnetic Materials and Metamaterials* (METAMORPHOSE VI, the International Metamaterials Society) and is an Elected Member of the Board of Directors of the same association from three consecutive terms (2014-now). In 2022 he has been appointed as member of the IEEE *APS Award committee* by the IEEE APS Society. Davide Ramaccia serves as an Associate Editor for the *IEEE Access* (2019-now), a Scientific Moderator for *IEEE TechRxiv* (2019-now), a Technical Reviewer of the major international journals related to electromagnetic field theory and metamaterials. He was also as Guest co-editor of three special issues on metamaterials and metasurfaces.

Since 2015, he serves as a member of the Steering Committee of the *International Congress on Advanced Electromagnetic Materials in Microwaves and Optics – Metamaterials Congress*. He has been General Chair and Local organizer of the 39th and 42nd *EUPROMETA* doctoral school on metamaterials held in Rome, Italy, in 2019 and 2021, respectively. He has been Technical Program Coordinator (Track "Electromagnetics and Materials") for the *2016 IEEE Antennas and Propagation Symposium.* He is member of the Technical Program Committee of the *International congress on Laser science and photonics applications - CLEO 2022.* He has also been elected as a Secretary of the Project Management Board of the *H2020 CSA project NANOARCHITECTRONICS* (2017–2018).

Davide Ramaccia was the recipient of a number of awards and recognitions, including *The Electromagnetics Academy Young Scientist Award* (2019) seven *Outstanding Reviewer Award by the IEEE Transactions on Antennas and Propagation* (2013-2021), the *IET prizes for the best poster on microwave metamaterials* (2013) and *IET Award for the Best Poster on the Metamaterial Application in Antenna Field* (2011).

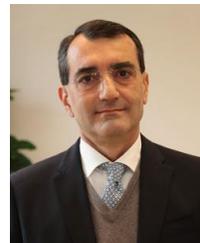

**ALESSANDRO TOSCANO** (M'91–SM'11) (Capua, 1964) graduated in Electronic Engineering from Sapienza University of Rome in 1988 and he received his PhD in 1993. Since 2011, he has been Full Professor of Electromagnetic Fields at the Engineering Department of Roma Tre University. He carries out an intense academic and scientific activity, both nationally and internationally.

From April 2013 to January 2018, he was a member of Roma Tre University Academic Senate. From October 2016 to October 2018, he is a member of the National Commission which enables National Scientific Qualifications to Full and Associate Professors in the tender sector 09/F1 – Electromagnetic fields. Since 23rd January 2018 he has been Vice-Rector for Innovation and Technology Transfer.

In addition to his commitment in organizing scientific events, he also carries out an intense editorial activity as a member of the review committees of major international journals and conferences in the field of applied electromagnetics. He has held numerous invited lectures at universities, public and private research institutions, national and international companies about artificial electromagnetic materials, metamaterials and their applications. He actively participated in founding the international association on metamaterials Virtual Institute for Advanced Electromagnetic Materials – METAMORPHOSE, VI. He coordinates and participates in several research projects and contracts funded by national and international public and private research institutions and industries.

Alessandro Toscano's scientific research has as ultimate objective the conceiving, designing, and manufacturing of innovative electromagnetic components with a high technological content that show enhanced performance compared to those obtained with traditional technologies and that respond to the need for environment and human health protection. His research activities are focused on three fields: metamaterials and unconventional materials, in collaboration with Professor A. Alù's group at The University of Texas at Austin, USA, research and development of electromagnetic cloaking devices and their applications (First place winner of the Leonardo Group Innovation Award for the research project entitled: 'Metamaterials and electromagnetic invisibility') and the research and manufacturing of innovative antenna systems and miniaturized components (first place winner of the Leonardo Group Innovation Award for the research project entitled: "Use of metamaterials for miniaturization of components" – MiniMETRIS).

He is the author of more than one hundred publications in international journals indexed ISI or Scopus; of these on a worldwide scale, three are in the first 0.1 percentile, five in the first 1 percentile and twenty-five in the first 5 percentile in terms of number of quotations and journal quality.

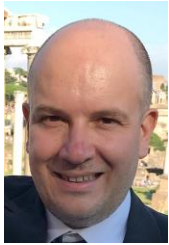

**FILIBERTO BILOTTI** (S'97–M'02–SM'06–F'17) received the Laurea and Ph.D. degrees in electronic engineering from ROMA TRE University, Rome, Italy, in 1998 and 2002, respectively. Since 2002, he has been with the Faculty of Engineering (2002-2012) and, then, with the Department of Engineering (2013-now), at ROMA TRE University, where he serves as a Full Professor of electromagnetic field theory (2014-now) and the *Director of the Antennas and Metamaterials Research Laboratory* (2012-present).

His main research contributions are in the analysis and design of microwave antennas and arrays, analytical modelling of artificial electromagnetic materials, metamaterials, and metasurfaces, including their applications at both microwave and optical frequencies. In the last ten years, Filiberto Bilotti's main research interests have been focused on the analysis and design of cloaking metasurfaces for antenna systems, on the modelling and applications of (space and) time-varying metasurfaces, on the topological-based design of antennas supporting structured field, on the modelling, design, and implementation of non-linear and reconfigurable metasurfaces, on the concept of meta-gratings and related applications in optics and at microwaves, on the modelling and applications of optical metasurfaces. The research activities developed in the last 20 years (1999-2019) has resulted in more than 500 papers in international journals, conference proceedings, book chapters, and 3 patents.

Prof. Bilotti has been serving the scientific community, by playing leading roles in the management of scientific societies, in the editorial board of international journals, and in the organization of conferences and courses. In particular, he was a founding member of the *Virtual Institute for Artificial Electromagnetic Materials and Metamaterials* – METAMORPHOSE VI in 2007. He was elected as a member of the Board of Directors of the same society for two terms (2007-2013) and as the President for two terms (2013-2019).

Currently, he serves the METAMORPHOSE VI as the Vice President and the Executive Director (2019-now).

Filiberto Bilotti served as an Associate Editor for the *IEEE Transactions on Antennas and Propagation* (2013-2017) and the journal *Metamaterials* (2007-2013) and as a member of the Editorial Board of the *International Journal on RF and Microwave Computer-Aided Engineering* (2009-2015), *Nature Scientific Reports* (2013-2016), and *EPJ Applied Metamaterials* (2013-now). He was also the Guest Editor of 5 special issues in international journals.

He hosted in 2007 the inaugural edition of the *International Congress on Advanced Electromagnetic Materials in Microwaves and Optics – Metamaterials Congress*, served as the Chair of the Steering Committee of the same conference for 8 editions (2008-2014, 2019), and was elected as the General Chair of the *Metamaterials Congress* for the period 2015-2018. Filiberto Bilotti was also the General Chair of the *Second International Workshop on Metamaterials-by-Design Theory, Methods, and Applications to Communications and Sensing* (2016) and has been serving as the chair or a member of the technical program, steering, and organizing committee of the main national and international conferences in the field of applied electromagnetics.

Prof. Bilotti was the recipient of a number of awards and recognitions, including the elevation to the *IEEE Fellow* grade for contributions to metamaterials for electromagnetic and antenna applications (2017), *outstanding Associate Editor of the IEEE Transactions on Antennas and Propagation* (2016), *NATO SET Panel Excellence Award* (2016), *Finmeccanica Group Innovation Prize* (2014), *Finmeccanica Corporate Innovation Prize* (2014), IET Best Poster Paper Award (Metamaterials 2013 and Metamaterials 2011), Raj Mittra Travel Grant Senior Researcher Award (2007).